\documentclass[prb,twocolumn,superscriptaddress,a4paper,twoside]{revtex4}

\usepackage{amsmath}
\usepackage{amssymb}
\usepackage{todonotes}
\usepackage{nicefrac}
\usepackage{graphics}
\usepackage{color}
\usepackage{graphicx}
\usepackage{verbatim}
\usepackage{float}

\makeatletter
\newcommand*{\rom}
[1]{\expandafter\@slowromancap\romannumeral #1@}
\makeatother


\begin{document}

\title{Time-dependent exchange-correlation potential in lieu of self-energy}

\author{F. Aryasetiawan}
\affiliation{Department of Physics, Division of Mathematical Physics, Lund University, Professorsgatan 1, 223 63, Lund, Sweden}

\begin{abstract}
It is shown that the equation of motion of the one-particle Green function of
an interacting many-electron system is governed by a multiplicative
time-dependent exchange-correlation potential, which is the Coulomb potential
of a time-dependent exchange-correlation hole. This exchange-correlation hole
fulfills a sum rule, a generalization of the well-known sum rule of the static
exchange-correlation hole. It is envisaged that the proposed formalism may
provide an alternative route for calculating the Green function by finding a
suitable approximation for the exchange-correlation hole or potential based
on, e.g., a local-density approximation.
\end{abstract}

\maketitle

\section{Introduction}

The one-particle Green function of an interacting many-particle system is of
utmost importance in condensed-matter physics and other branches of physics
such as molecular physics and nuclear physics. Essential physical properties,
notably the ground-state expectation value of a
single-particle operator, the total energy, and the particle addition and
removal spectra can be extracted from the one-particle Green function,
hereafter referred to simply as the Green function. Experimental photoemission
and inverse photoemission spectra, which provide invaluable information about
the electronic structure of the system, can be directly compared with the
spectra extracted from the Green function, under the so-called sudden
approximation and neglecting the matrix-element effects. Experimental data
from transport measurements and many other experimental observations can also
be related to the Green function. A lot of efforts have therefore been
expended on developing methods and techniques for calculating the Green
function, from many-body perturbation theory \cite{fetter-walecka}
to methods employing path-integral techniques \cite{negele}.

The zero-temperature time-ordered Green function is defined according to
\cite{fetter-walecka}
\begin{equation}
iG(rt,r^{\prime}t^{\prime})=\langle  T[\hat{\psi
}(rt\mathbf{)}\hat{\psi}^{\dag}(r^{\prime}t^{\prime})]\rangle ,
\end{equation}
where $r=(\mathbf{r,}\sigma)$ is a combined label of space and spin variables,
$\hat{\psi}(rt\mathbf{)}$ is the field operator in the Heisenberg picture, $T$
is the time-ordering symbol, and the expectation value is taken with
respect to the ground-state. The
many-electron Hamiltonian defining the Heisenberg operator is given by
\begin{align}
\hat{H}  &  =\int dr\text{ }\hat{\psi}^{\dag}(r)h_{0}(r)\hat{\psi
}(r)\nonumber\\
&  +\frac{1}{2}\int drdr^{\prime}\text{ }\hat{\psi}^{\dag}(r)\hat{\psi}^{\dag
}(r^{\prime})v(r-r^{\prime})\hat{\psi}(r^{\prime})\hat{\psi}(r),
\label{H}
\end{align}
where $h_{0}=-\frac{1}{2}\nabla^{2}+V_{ext}(r)$ and $v(r-r^{\prime
})=1/|\mathbf{r-r}^{\prime}|$. In our notation, $\int dr=\sum_{\sigma}\int
d^{3}r$ and atomic unit is used throughout, in which the Bohr radius $a_{0}$,
the electron mass $m_{e}$, the electronic charge $e$, and $\hbar$ are set to unity.

Since the Hamiltonian is time independent it is convenient to set $t^{\prime
}=0$. The Green function fulfills the following equation of motion:
\begin{align}
&  \left(  i\frac{\partial}{\partial t}-h_{0}(r\mathbf{)}\right)
G(r,r^{\prime};t)\nonumber\\
&  +i\int dr^{\prime\prime}v(r-r^{\prime\prime})\langle
T[\hat{\rho}(r^{\prime\prime}t)\hat{\psi}(rt\mathbf{)}%
\hat{\psi}^{\dag}(r^{\prime})]\rangle \nonumber\\
&  =\delta(r-r^{\prime})\delta(t). \label{EOM-G1}%
\end{align}
where $\hat{\rho}$ is the density operator. The interaction term contains a
special case of the two-particle Green function:
\begin{equation}
G^{(2)}(r,r^{\prime},r^{\prime\prime};t)=\langle 
T[\hat{\rho}(r^{\prime\prime}t)\hat{\psi}(rt\mathbf{)}\hat{\psi}^{\dag
}(r^{\prime})]\rangle . \label{G2-0}%
\end{equation}
The traditional approach is to introduce a self-energy $\Sigma$ as follows:
\begin{align}
-  &  i\int dr^{\prime\prime}v(r-r^{\prime\prime})G^{(2)}(r,r^{\prime
},r^{\prime\prime};t)-V_{H}(r)G(r,r^{\prime};t)\nonumber\\
&  =\int dr^{\prime\prime}dt^{\prime\prime}\Sigma(r,r^{\prime\prime
};t-t^{\prime\prime})G(r^{\prime\prime},r^{\prime};t^{\prime\prime}),
\label{Sigma}%
\end{align}
where $V_{H}$ is the Hartree potential subtracted from $G^{(2)}$. Thus, the
self-energy embodies the effects of exchange and correlations and a central
quantity in Green function theory. The self-energy is a well-established
framework for calculating the Green function but it acts on the Green function
as a convolution in space and time and as such it is difficult to visualize
its meaning in a clear physical picture. Moreover, from the Dyson equation,
$G=G_{0}+G_{0}\Sigma G$, it can be seen that $\Sigma$ is an auxiliary quantity
since it depends on the choice of the starting reference Green function
$G_{0}$.

In this paper, a completely different route is proposed to calculate the Green
function. It is shown that the equation of motion of the Green function can be
reformulated so that it is governed by a time-dependent exchange-correlation
potential that acts multiplicatively on the Green function. This
exchange-correlation potential arises naturally from the Coulomb potential of
the time-dependent exchange correlation hole, which fulfills a sum rule. The
static and equal-space limit $r^{\prime}\rightarrow r$ of this
exchange-correlation hole reduces to the well-known static
exchange-correlation hole in the formal expression for the
exchange-correlation energy. The focus is then shifted to finding an accurate
approximation for the exchange-correlation hole or potential, in the spirit of
density functional theory \cite{hohenberg1964,kohn1965,jones1989,becke2014}.

\section{Theory}

\subsection{Time-dependent exchange-correlation hole}

Writing out the time ordering of the two-particle Green function in Eq.
(\ref{G2-0}) and using the commutation
\begin{equation}
\lbrack\hat{\psi}(rt),\hat{\rho}(r^{\prime\prime}t)]=\delta(r-r^{\prime\prime
})\hat{\psi}(rt).
\end{equation}
yields
\begin{align}
&  G^{(2)}(r,r^{\prime},r^{\prime\prime};t)=\langle \hat{\rho
}(r^{\prime\prime}t)\hat{\psi}(rt)\hat{\psi}^{\dag}(r^{\prime})\rangle \theta(t)\nonumber\\
&  \text{ \ \ \ \ \ \ \ \ \ \ \ \ \ \ \ \ \ \ \ }-\langle \hat{\psi}^{\dag}(r^{\prime})\hat{\psi}(rt)\hat{\rho}(r^{\prime\prime}%
t)\rangle \theta(-t)\nonumber\\
&  \text{ \ \ \ \ \ \ \ \ \ \ \ \ \ \ }+\delta(r-r^{\prime\prime})\langle
\hat{\psi}^{\dag}(r^{\prime})\hat{\psi}(rt)\rangle
\theta(-t). \label{G20}%
\end{align}

Consider now integrating $G^{(2)}$ over the variable $r^{\prime\prime}$. It
should first be noted that for any state $\Psi$ containing $N$ electrons,
\begin{equation}
\int dr^{\prime\prime}\hat{\rho}(r^{\prime\prime})\left\vert \Psi\right\rangle
=\hat{N}\left\vert \Psi\right\rangle =N\left\vert \Psi\right\rangle ,
\end{equation}
where $\hat{N}$ counts the number of electrons in the system. Using $G^{(2)}$
in Eq, (\ref{G20}) one finds for $t<0$
\begin{equation}
\int dr^{\prime\prime}G^{(2)}(r,r^{\prime},r^{\prime\prime};t)=iG(r,r^{\prime
};t)\left(  N-1\right)  . \label{intG2}%
\end{equation}
$G^{(2)}$ can be naturally factorized as follows:
\begin{equation}
G^{(2)}(r,r^{\prime},r^{\prime\prime};t)=iG(r,r^{\prime};t)g(r,r^{\prime
},r^{\prime\prime};t)\rho(r^{\prime\prime}), \label{gh}%
\end{equation}
which defines the correlation function $g(r,r^{\prime},r^{\prime\prime};t)$,
and after substitution into Eq. (\ref{intG2}) one arrives at the sum rule
\begin{equation}
\int dr^{\prime\prime}[g(r,r^{\prime},r^{\prime\prime};t)-1]\rho
(r^{\prime\prime})=-1. \label{sum-rule0}%
\end{equation}
This sum rule is valid for any $r,r^{\prime}$, and $t<0$ and the integrand may
be interpreted as the time-dependent exchange-correlation hole:
\begin{equation}
\rho_{xc}(r,r^{\prime},r^{\prime\prime};t)=[g(r,r^{\prime},r^{\prime\prime
};t)-1]\rho(r^{\prime\prime}), \label{xchole}%
\end{equation}
a generalization of the static exchange-correlation hole, first introduced by
Slater for the exchange part
\cite{slater1951}. Since $\rho_{xc}(r,r^{\prime},r^{\prime\prime};t)$
is in general complex, the sum rule implies that the imaginary part integrates
to zero.

It is interesting to observe that for $t>0$ (addition of an electron), the
exchange-correlation hole integrates to zero as can be seen from Eq.
(\ref{G20}). This result may be understood by recognizing that the added
electron is not part of the electron density that generates the Hartree
potential so that there is no self-interaction corresponding to the last term
of $G^{(2)}$ in Eq. (\ref{G20}). For $\sigma^{\prime\prime}\neq\sigma$, the
exchange-correlation hole integrates to zero as may be seen from the presence
of $\delta(r-r^{\prime\prime})=\delta(\mathbf{r-r}^{\prime})\delta
_{\sigma\sigma^{\prime\prime}}$ in Eq. (\ref{G20}). If one were to decompose
$\rho_{xc}$ into the exchange and the correlation holes, it is the exchange
hole that would integrate to $-1$ whereas the correlation hole would integrate
to zero. The sum rule can then be summarized as follows:
\begin{align}
\int d^{3}r^{\prime\prime}\rho_{xc}(r,r^{\prime},r^{\prime\prime};t)
&=-\delta_{\sigma\sigma^{\prime\prime}} \theta(-t),\label{sum-rule}
\end{align}
This sum rule may be viewed as the generalization of the well-known sum rule
for the static exchange-correlation hole appearing in the formally exact
expression for the exchange-correlation energy \cite{jones1989,becke2014},
originating from Slater's sum rule for exchange hole only \cite{slater1951}.
The static
exchange-correlation hole corresponds to the limit $t\rightarrow0^{-}$ and
$r^{\prime}\rightarrow r.$

\subsection{Time-dependent exchange-correlation potential}

Rearranging Eq. (\ref{gh}) yields
\begin{align}
&  G^{(2)}(r,r^{\prime},r^{\prime\prime};t)\nonumber\\
&  =iG(r,r^{\prime};t)\rho(r^{\prime\prime})+iG(r,r^{\prime};t)\rho
_{xc}(r,r^{\prime},r^{\prime\prime};t), \label{Gxc}%
\end{align}
in which the first term on the right-hand side generates the Hartree term.
When the above $G^{(2)}$ is substituted into the equation of motion of $G$ in
Eq. (\ref{EOM-G1}), it generates the time-dependent exchange-correlation
potential as a function of $r,r^{\prime}$ and $t$:
\begin{equation}
V_{xc}(r,r^{\prime};t)=\int dr^{\prime\prime}v(r-r^{\prime\prime})\rho
_{xc}(r,r^{\prime},r^{\prime\prime};t),
\label{Vxc}
\end{equation}
acting on $G$ in a multiplicative fashion, in contrast to the self-energy
which acts on $G$ as a convolution in space-time as in Eq. (\ref{Sigma}). The
equation of motion of $G$ becomes
\begin{equation}
\left(  i\frac{\partial}{\partial t}-h(r\mathbf{)}-V_{xc}(r,r^{\prime
};t)\right)  G(r,r^{\prime};t)=\delta(r-r^{\prime})\delta(t), \label{EOM}%
\end{equation}
where $h=h_{0}+V_{H}$. This equation has a local character in the sense that
the potential is multiplicative in both space and time. Apart from the source
term on the right-hand side, for a given $r^{\prime}$ it is just like a
one-particle time-dependent Schr\"{o}dinger equation in the presence of a
time-dependent field $V_{xc}$. The effects of exchange and correlations are
now embodied in the time-dependent exchange-correlation potential $V_{xc}$,
which is in general non-Hermitian. Eq. (\ref{EOM}) furnishes us with a
different picture from that of the conventional self-energy formulation and
offers a simple physical interpretation for the propagation of an added
electron or hole, which is governed by, in addition to the external field and
the Hartree potential, the time-dependent exchange-correlation potential. This
exchange-correlation potential is simply the Coulomb potential of the
time-dependent exchange-correlation hole.

The corresponding Dyson-like equation for $G$ can be readily written down:
\begin{align}
    &G(r,r';t) = G_H(r,r';t) 
    \nonumber\\
    &+ \int dr''dt' G_H(r,r'';t-t') 
    V_{xc}(r'',r^{\prime};t')G(r'',r^{\prime};t'),
\end{align}
where $G_H$ is the Hartree Green function, 
and the relationship between $V_{xc}$ and $\Sigma$ in space-time is given by
\begin{equation}
V_{xc}(r,r^{\prime};t)G(r,r^{\prime};t)=\int dr^{\prime\prime}dt^{\prime
}\Sigma(r,r^{\prime\prime};t-t^{\prime})G(r^{\prime\prime},r^{\prime
};t^{\prime}). \label{VxcSigma}
\end{equation}

Expressed in a set of base orbitals $\{\varphi_{i}\}$, the equation
of motion in Eq. (\ref{EOM}) takes the
form
\begin{equation}
i\frac{\partial}{\partial t}G_{ij}(t)-\sum_{k}h_{ik}G_{kj}(t)-\sum
_{kl}V_{ik,lj}^{xc}(t)G_{kl}(t)=\delta_{ij}\delta(t),
\end{equation}
where $G_{ij}$ and $h_{ik}$ are the matrix elements of $G$ and $h$ in the
orbitals and
\begin{equation}
V_{ik,lj}^{xc}(t)=\int d^{3}rd^{3}r^{\prime}\text{ }\varphi_{i}^{\ast}(r)
\varphi_{k}(r)
V_{xc}(r,r^{\prime};t)\varphi_{l}^{\ast}(r^{\prime})\varphi_{j}(r^{\prime}),
\end{equation}
exhibiting the two-particle character of $V_{xc}$, a Bosonic quantity, in
contrast to the self-energy which is Fermionic.

Much is known about the static exchange-correlation hole \cite{becke2014},
which may provide a starting point for finding a good approximation for the
time-dependent one. Approximating $V_{xc}$ via the
exchange-correlation hole $\rho_{xc}$, which is a physically motivated entity,
may be more advantageous than following the arduous route of finding a good
approximation for the self-energy by means of many-body perturbation theory or
path integral approach. The proposed formalism has a certain proximity to
time-dependent density functional theory \cite{runge1984,maitra2016}. 
It should be noted, however, that
the time dependence of $V_{xc}$ does not arise from the presence of a
time-dependent external field, but rather due to the dynamics of the Green
function. The addition or removal of an electron causes the system to evolve
in a non-trivial way due to the Coulomb interaction among the electrons.

\subsection{Local-density approximation}

It can be anticipated that $V_{xc}$ is a relatively smooth function since it
is the Coulomb potential of a charge distribution that integrates to $-1$ or
$0$. Following Gunnarsson and Lundqvist \cite{gunnarsson1976,jones1989}, 
it is readily seen by
making a change of variable $\mathbf{R}=\mathbf{r}-\mathbf{r}^{\prime\prime}$
that, due to the form of the Coulomb interaction, only the spherical average
of the exchange-correlation hole in the variable $\mathbf{R}$ is needed to
determine $V_{xc}$:
\begin{equation}
V_{xc}(r,r^{\prime};t)=\sum_{\sigma_{R}}\int dRR\text{ }\overline{\rho}%
_{xc}(r,r^{\prime},R;t),
\label{Vxc1mom}
\end{equation}
where $\overline{\rho}_{xc}(r,r^{\prime},R;t)$ depends only on the radial
distance $R$ with respect to $\mathbf{r}$,
\begin{equation}
\overline{\rho}_{xc}(r,r^{\prime},R;t)=\int d\Omega_{R}\rho_{xc}(r,r^{\prime
},\mathbf{r-R};t),
\end{equation}
implying that the fine spatial details of the exchange-correlation hole may
not be important as illustrated vividly for the case of the static
exchange-correlation hole in some light atoms \cite{gunnarsson1979,jones1989}. 
According to Eq. (\ref{Vxc1mom}), $V_{xc}$ is the
first radial moment of the spherically averaged $\rho_{xc}$.

It can also be
seen that $\mathbf{r}$ may be thought of as the center of the
exchange-correlation hole whereas $r^{\prime}$ may be treated as a parameter
representing the spatial origin of the created electron. One could imagine
that the exchange-correlation hole moves with the added hole or electron as a
function of time.
From Eq. (\ref{G20}) it is quite evident that when $r^{\prime\prime}=r$,
$G^{(2)}=g=0$ and hence
\begin{equation}
    \rho_{xc}(r,r^{\prime},r;t)=-\rho(r)
\end{equation}
for \emph{any}
$r,$ $r^{\prime},$ and $t$. Following Slater's argument \cite{slater1968}, it
can be concluded that the exchange-correlation potential behaves approximately
as $\rho^{1/3}$. It may be envisaged that a local-density approximation for
$V_{xc}$ as in density functional theory can be developed and its
time-dependence can be constructed from knowledge of the time-dependent
exchange-correlation hole of the electron gas and some generic model systems,
depending on the correlation strength. 

The correlation function of the
homogeneous electron gas with density $\Bar{\rho}$ can be written as follows:
\begin{equation}
g^{HEG}_{\sigma\sigma''}(\Bar{\rho},R',R'',\theta;t),
\end{equation}
where $R'=|\mathbf{r-r'}|$, $R''=|\mathbf{r-r''}|$, $\theta$ is the angle
between $\mathbf{R}'$ and $\mathbf{R}''$, and the spin dependence has been written
explicitly.
A simple local-density approximation for the exchange-correlation hole could be
\begin{align}
&\rho_{xc}^{LDA}(r,r',r'';t) 
\nonumber\\
&= \left[g^{HEG}_{\sigma\sigma''}(\rho(r),R',R'',\theta;t)-1\right]\rho(r'').
\label{xc-hole_LDA}
\end{align}

A more sophisticated approximation would be to employ the weighted-density approximation
\cite{alonso1978,gunnarsson1979}, in which the density dependence associated with
the variable $r'$ is taken into account. 
On the other hand, applying the local-density approximation directly on $V_{xc}$,
\begin{equation}
V_{xc}^{LDA}(r,r';t) = V_{xc}^{HEG}(\rho(r),R;t),
\end{equation}
where $R=|\mathbf{r-r'}|$, may be too crude since information encoded in $\rho(r'')$ is lost.
It would be interesting to compare this approximation with
the local-density approximation for the self-energy based on the homogeneous electron gas
proposed by Sham and Kohn many years ago \cite{sham1966}. One may speculate that
a local-density approximation on $V_{xc}$ is more favorable than on $\Sigma$
since $V_{xc}$ acts locally on $G$. Also, in contrast to $\Sigma$, which is Fermionic, 
$V_{xc}$ is a Bosonic object and as such it may be easier to approximate in terms of
the density, which is a Bosonic quantity.

\subsection{Connection with Kohn-Sham $V_{xc}$}

If $V_{xc}(r,r';t)$ is approximated by the static 
Kohn-Sham exchange-correlation potential,
$V^{KS}_{xc}(r)$, then the Green function reduces to the Kohn-Sham
non-interacting Green function, whose diagonal component will by construction yield
the exact density. 
However, $V^{KS}_{xc}(r)$ is not necessarily the same as
$V_{xc}(r,r;0^-)$.
From the equations of motion of $G$ and $G^{KS}$ one finds
\begin{align}
    &\left(  i\frac{\partial}{\partial t}-h(r\mathbf{)}
    -V_{xc}(r,r^{\prime};t)\right)  [G(r,r^{\prime};t)-G^{KS}(r,r^{\prime};t)]
    \nonumber\\
    &=[V_{xc}(r,r^{\prime};t)-V^{KS}_{xc}(r)] G^{KS}(r,r^{\prime};t).
\end{align}
Evaluating the equation at $r'\rightarrow r$ and $t\rightarrow 0^-$ and making use of the fact
that both $G$ and $G^{KS}$ give the same density yields the relationship between
$V_{xc}(r,r;0^-)$ and $V^{KS}_{xc}(r)$:
\begin{align}
    &\lim_{r'\rightarrow r,t\rightarrow 0^-}\left(  \frac{\partial}{\partial t}-\frac{i}{2}\nabla^2\right)  [G(r,r^{\prime};t)-G^{KS}(r,r^{\prime};t)]
    \nonumber\\
    &=[V_{xc}(r,r;0^-)-V^{KS}_{xc}(r)] \rho(r).
    \label{id0}
\end{align}
There is no obvious reason that the left-hand side vanishes so that in general 
$V_{xc}(r,r;0^-) \neq V^{KS}_{xc}(r)$. 
When integrated over $r$,
the first term on the left-hand side involving time derivative is the difference
in the first moment of the occupied densities of states,
whereas the second term is the difference in the kinetic energies, which is contained in
the Kohn-Sham $E_{xc}$.

One also notes that
\begin{equation}
g(r,r,r'';t=0^-) \neq g^{KS}(r,r''),
\end{equation}
since $g^{KS}$ is defined as a coupling-constant integration of $g_\lambda$, which is the
static correlation function corresponding to a scaled Coulomb interaction, $\lambda v$,
yielding the
exact ground-state density independent of $\lambda$ \cite{gunnarsson1976,perdew1975}.

The exchange-correlation energy in the Kohn-Sham scheme is given by
\begin{equation}
    E^{KS}_{xc}= \frac{1}{2}\int dr dr' \rho(r) v(r-r')\rho_{xc}^{KS}(r,r'),
\end{equation}
where
\begin{equation}
    \rho_{xc}^{KS}(r,r') = [g^{KS}(r,r')-1]\rho(r')
\end{equation}
is the static Kohn-Sham exchange-correlation hole.
It can then be seen that the Kohn-Sham exchange-correlation potential is given by
\begin{align}
    V^{KS}_{xc}(r)&=\int dr' v(r-r')\rho_{xc}^{KS}(r,r') 
    \nonumber\\
    &+\frac{1}{2}\int dr' dr'' \rho(r') v(r'-r'')
    \frac{\delta g^{KS}(r',r'')}{\delta \rho(r)}\rho(r'').
\end{align}
Thus in addition to the Coulomb potential of the exchange-correlation hole, which is the counterpart
of $V_{xc}$ in Eq. (\ref{Vxc}), there is an additional
contribution arising from the dependence of the distribution function $g^{KS}$ on the density.


\subsection{Kohn-Sham scheme for unoccupied states}

Since $V_{xc}(0^-)\neq V_{xc}(0^+)$, there will in general be a discontinuity at $t=0$.
This suggests that in the Kohn-Sham scheme, two exchange-correlation potentials are needed,
one for the occupied states and another for the unoccupied ones.
For the occupied states, the standard Kohn-Sham equation applies while for the unoccupied states
one has
\begin{equation}
    \left( -\frac{1}{2} \nabla^2 + V_{ext}+V_{H}+V^+_{xc}\right) \phi_{kn}
    =\epsilon_k \phi_{kn},
    \label{KS+}
\end{equation}
where $V^+_{xc}$ is the exchange-correlation potential corresponding to an
exchange-correlation hole that integrates to zero rather than to one. 
After solving the standard Kohn-Sham equation, the unoccupied states are used to diagonalize
Eq. (\ref{KS+}), which ensures that all states will be orthogonal.
A local density approximation for the exchange-correlation hole corresponding to
$V^+_{xc}$ can be taken to be the one
in Eq. (\ref{xc-hole_LDA}) with $r'=r$ and $t=0^+$.

Since the exchange-correlation hole corresponding to $V^+_{xc}$ integrates to zero, it should be weaker than the one for the occupied states for weakly or moderately correlated systems.
This implies
that the unoccupied states would be pushed up, 
correcting the well-known underestimation of band gaps in semiconductors and 
insulators. For metals, however, one may reason that since the band dispersion is smooth
across the Fermi surface, the discontinuity should tend to vanish.

As an approximate
$V^+_{xc}$ it may be reasonable to use only the correlation part of
the standard $V_{xc}$.
A simple correction would be to use first-order perturbation theory:
\begin{equation}
    \varepsilon_{kn}^+ = \varepsilon_{kn} + \langle \phi_{kn} |V^+_{xc}-V_{xc}|\phi_{kn}\rangle.
\end{equation}
An even simpler correction would be to ignore entirely $V^+_{xc}$, which may lead, however,
to an overestimation of the gap since $V^+_{xc}$ is likely to be negative.

\subsection{Extension to temperature-dependent Green function}

The temperature-dependent Green function is defined in a similar way as for the
zero-temperature one with the ground-state expectation value replaced by the thermal average:
\begin{align}
    iG_\beta(rt,r't') &= \frac{1}{Z} \mathrm{Tr}\left\{ e^{-\beta\hat{K}} T [ \hat{\psi}(rt)   \hat{\psi}^\dagger(r't')]  \right\},
\end{align}
where $\hat{K}= \hat{H} - \mu \hat{N}$, $\mu$ is the chemical potential, 
$Z=\mathrm{Tr}\:e^{-\beta\hat{K}}$ is the partition function, and 
$\beta=1/(k_B T)$ with $k_B$ being 
the Boltzmann constant and $T$ the temperature. The Heisenberg operators is defined as
\begin{align}
    \hat{\psi}(rt) = e^{i\hat{K}t} \hat{\psi}(r) e^{-i\hat{K}t}.
\end{align}

At this stage, one traditionally goes over to imaginary time yielding the Matsubara 
Green function. The reason is that Wick's theorem is no longer
convenient to use if one stays along the real-time axis due to the presence of
the thermal factor $e^{-\beta\hat{K}}$. The proposed formalism, however, 
makes no use of Wick's theorem so that it is not necessary to work along the imaginary-time
axis. The ill-defined problem of analytic continuation to the real-time axis when calculating spectral functions associated with the Matsubara Green function is circumvented. 

The field operators are independent of temperature so that the equation of motion
of $G_\beta$ is the same as the one in Eq. (\ref{EOM-G1})
with the understanding that ground-state expectation value is now to be understood
as thermal average and $h_0$ is replaced with $h_0-\mu$. 
It is quite evident that
the sum rule and the equation of motion in Eq. (\ref{EOM}) still hold with $h$ replaced by $h-\mu$ and all quantities carry the temperature label $\beta$.
As in the zero-temperature case, the
main task is to find a good approximation for the temperature-dependent $\rho_{xc}$ or
$V_{xc}$. 

\subsection{Extension to non-equilibrium Green function}

For a non-equilibrium system, one must keep the time variable $t'$ but otherwise the
derivation is very similar. A time-dependent external field $\varphi(rt)$ is applied from
time $t=0$ and the system is assumed to be in the ground state for $t<0$, 
which would correspond to the pump-probe experiment. 

The equation of motion of the Green function is given by
\begin{align}
&\left(  i\frac{\partial}{\partial t}-h(rt)
-V_{xc}(rt,r^{\prime}t')\right)  
G(rt,r^{\prime}t')
\nonumber\\
&=\delta(r-r^{\prime})\delta(t-t'), \label{EOM-neq}%
\end{align}
where $h(rt)$ includes the time-dependent external field $\varphi(rt)$ and
\begin{equation}
V_{xc}(rt,r^{\prime}t') =\int dr'' v(r-r'')\rho_{xc}(rt,r't',r''),
\end{equation}
\begin{equation}
\rho_{xc}(rt,r't',r'') = [g(rt,r't',r'')-1]\rho(r'',t).
\end{equation}
The non-equilibrium correlation function $g$ is defined according to
\begin{equation}
G^{(2)}(rt,r^{\prime}t',r^{\prime\prime})=iG(rt,r^{\prime}t')g(rt,r^{\prime}t',r^{\prime\prime})
\rho(r^{\prime\prime},t), \label{gh-neq}%
\end{equation}
where
\begin{equation}
G^{(2)}(rt,r^{\prime}t',r^{\prime\prime})=
\langle 
T[\hat{\rho}(r^{\prime\prime}t)\hat{\psi}(rt)\hat{\psi}^{\dag
}(r^{\prime}t')]\rangle . \label{G2-neq}%
\end{equation}
The Heisenberg field operator $\hat{\psi}(rt)$ is defined more generally as
\begin{equation}
    \hat{\psi}(rt) = \hat{U}(0,t) \hat{\psi}(r) \hat{U}(t,0),
\end{equation}
where the time-evolution operator is given by
\begin{equation}
    \hat{U}(t,0) = T \exp{[-i\int_0^t dt' \hat{H}(t')]}.
\end{equation}
The time-dependent Hamiltonian $\hat{H}(t)$ is given by the many-electron Hamiltonian in 
Eq. (\ref{H}) including the time-dependent external field. $G^{(2)}$ and hence $V_{xc}$ and
$\rho_{xc}$ depend implicitly on $\varphi(rt)$.

The exchange-correlation hole fulfills the sum rule as in Eq. (\ref{sum-rule}):
\begin{equation}
\int d^3r'' \rho_{xc}(rt,r't',r'') = \delta_{\sigma\sigma''}\theta(t'-t),
\end{equation}
for any $r,t,r',t'$, independent of $\varphi(rt)$.

Since the formalism makes no use of Wick's theorem, it is not necessary to introduce the Keldysh or
the Kadanoff-Baym contours \cite{stefanucci2013}. 
Temperature dependence can be included as described in the previous subsection by simply replacing 
ground-state expectation value by thermal average.

A simple approximation could be to assume that
\begin{equation}
    g(rt,r't',r'') \approx g(r,r',r'';t-t')
\end{equation}
and make a local-density approximation as proposed in Eq. (\ref{xc-hole_LDA}).
The dependence on $\varphi(rt)$ enters through the time-dependent density.

\section{Examples}

To illustrate how the time-dependent exchange-correlation potential looks like
and as a proof of concept, the half-filled Hubbard dimer with total spin zero
is considered. Although it is very simple, it contains some of the essential
physics of correlated electrons and it has the great advantage of being
analytically solvable. 

Another exactly solvable Hamiltonian considered is a simplified Holstein model,
describing a core electron coupled to a set of Bosons
such as plasmons or phonons. This Hamiltonian is appropriate to
model solids in which the valence electrons are relatively delocalized, resembling electron gas.
The alkalis and $s$-$p$ semiconductors and insulators are examples of such systems.

\subsection{Hubbard dimer}

The Hamiltonian of the Hubbard dimer in standard
notation is given by
\begin{equation}
\hat{H}=-\Delta\sum_{i\neq j}\hat{c}_{i\sigma}^{\dag}\hat{c}_{j\sigma}%
+U\sum_{i}\hat{n}_{i\uparrow}\hat{n}_{i\downarrow}.
\end{equation}

\begin{figure}[htp]
\centering
\includegraphics[clip,width=0.9\columnwidth]{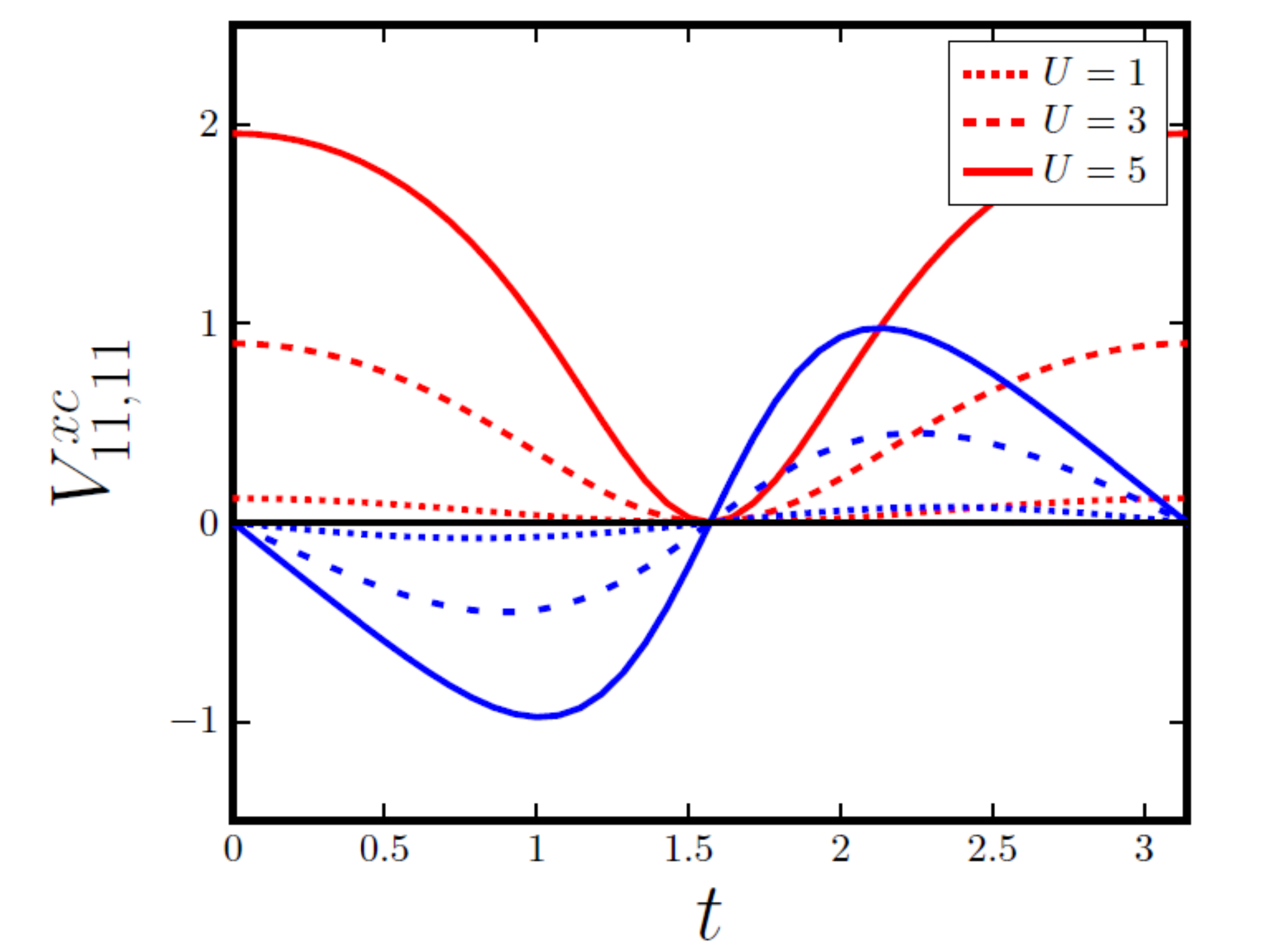}
\vfill
\hspace{1cm}
\includegraphics[clip,width=0.9\columnwidth]{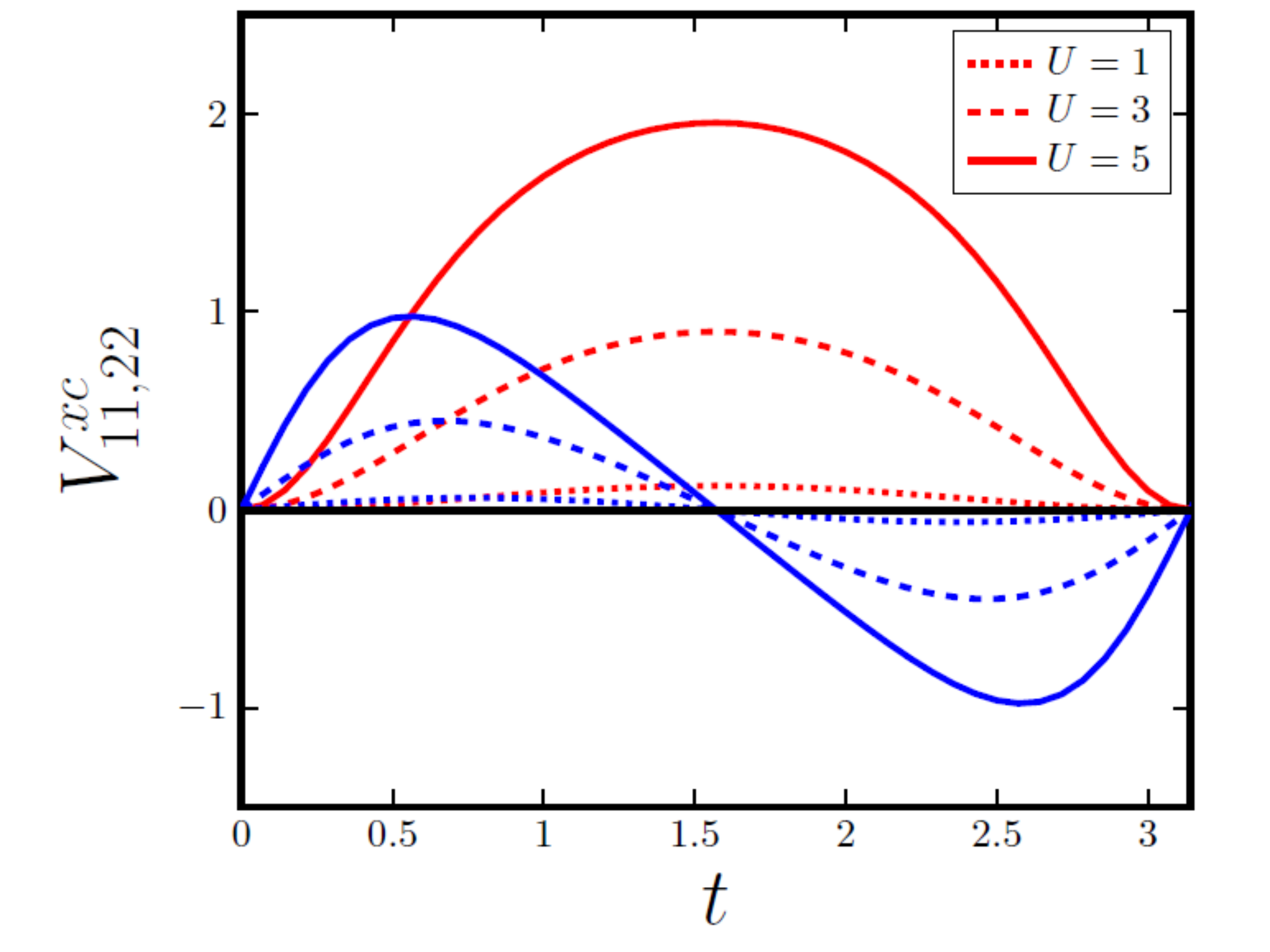}
\caption{The real (red) and imaginary (blue) parts of the exchange-correlation potentials 
$V^{xc}_{11,11}$ and $V^{xc}_{11,22}$ 
of the Hubbard dimer
as a function of time for $U=1,3,5$ with $\Delta=1$.
Due to the
particle-hole symmetry, $V_{xc}(-t)=-V_{xc}(t)$.}
\label{fig:Vxc}%
\end{figure}

\begin{figure}[htp]
\centering
\includegraphics[clip,width=0.9\columnwidth]{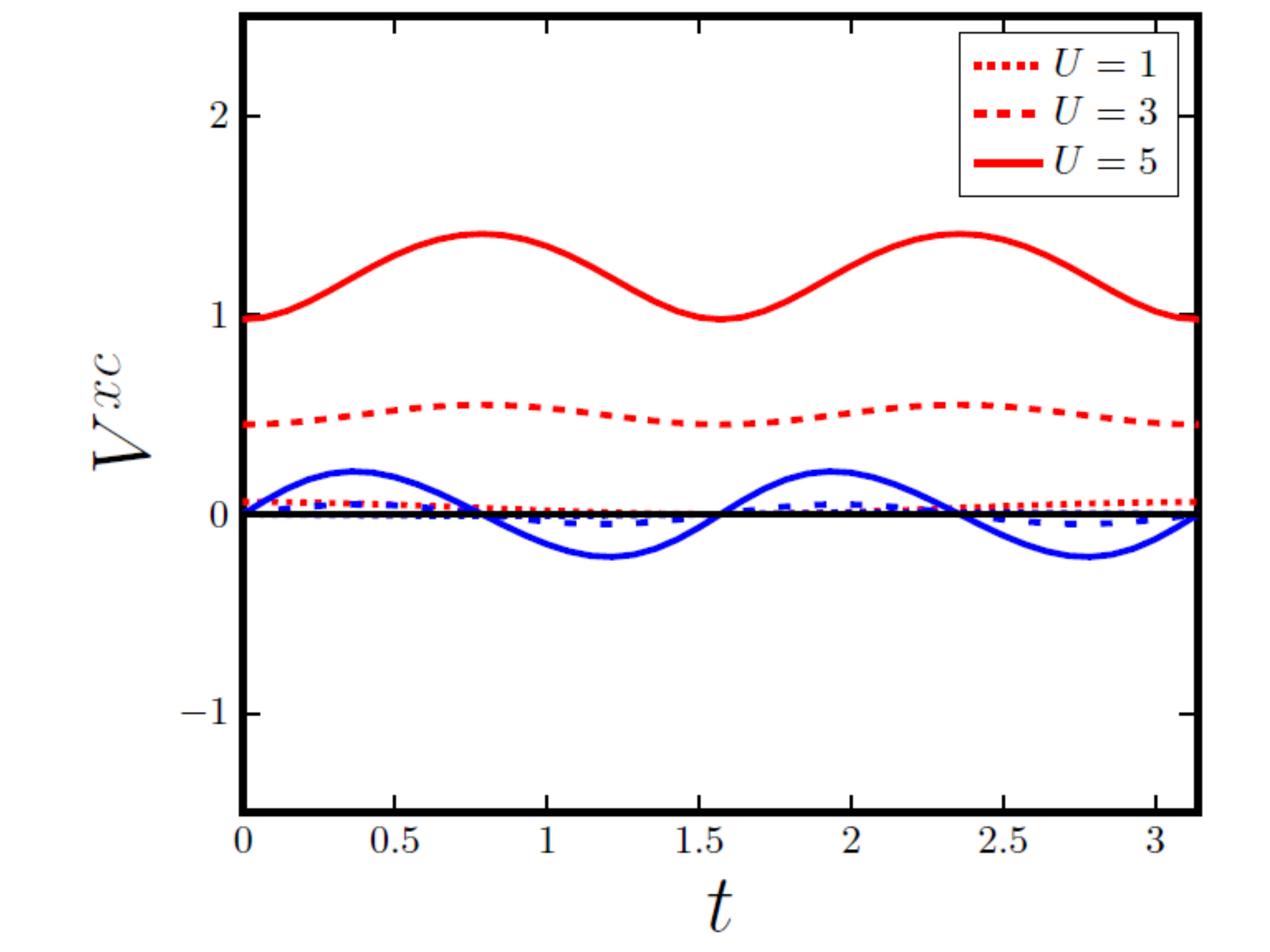}
\caption{The real (red) and imaginary (blue) parts of the exchange-correlation potential in
the bonding state $\phi_B=\frac{1}{\sqrt{2}}(\varphi_1+\varphi_2)$
or anti-bonding state $\phi_A=\frac{1}{\sqrt{2}}(\varphi_1-\varphi_2)$ as a function of time for $U=1,3,5$ with $\Delta=1$. 
Since only the components 
$V^{xc}_{11,11}=V^{xc}_{22,22}$ and $V^{xc}_{11,22}=V^{xc}_{22,11}$ exist, $V^{xc}_{BB,BB}=V^{xc}_{AA,AA}=\frac{1}{2}(V^{xc}_{11,11}+V^{xc}_{11,22})$.
The imaginary part of $V_{xc}$ for $U=1$ is practically invisible.}
\label{fig:VxcBA}%
\end{figure}

By calculating $G^{(2)}$ and using the relation in Eq. (\ref{Gxc}), $V_{xc}$
can be deduced. The results, shown in Fig. \ref{fig:Vxc} for $U=1,3,$ and $5$, are given by
\begin{equation}
V_{11,11}^{xc}(t>0)=\frac{U}{2}\frac{(1-x^{2})\left\{  1+\exp(-i2\Delta
t)\right\}  }{(1+x)^{2}+(1-x)^{2}\exp(-i2\Delta t)},
\end{equation}
\begin{equation}
V_{11,22}^{xc}(t>0)=\frac{U}{2}\frac{(1-x^{2})\left\{  1-\exp(-i2\Delta
t)\right\}  }{(1+x)^{2}-(1-x)^{2}\exp(-i2\Delta t)},%
\end{equation}
where
\begin{equation}
x=\frac{1}{4\Delta}\left(  \sqrt{U^{2}+16\Delta^{2}}-U\right)
\end{equation}
is the relative weight of double-occupancy configurations in the ground state
and
\begin{equation}
2\Delta=E_{1}^{-}-E_{0}^{-}=E_{1}^{+}-E_{0}^{+}>0,
\end{equation}
are the excitation energies of the $(N\pm1)$-systems. From symmetry,
$V_{22,22}^{xc}=V_{11,11}^{xc}$ and $V_{22,11}^{xc}=V_{11,22}^{xc}$, i.e.,
$V_{xc}$ is symmetric but it is not Hermitian since it is complex. Due to the
particle-hole symmetry, $V_{xc}(-t)=-V_{xc}(t)$. In Fig. \ref{fig:VxcBA}, 
the corresponding exchange-correlation potentials in the bonding or anti-bonding states
are shown. 

A number of generic conclusions can be drawn from this simple model.
Since $x$ depends on $U$, $V_{xc}$ is not simply proportional to the interaction $U$.
In the weakly
correlated limit corresponding to small $(1-x)$, it can be seen that the
dependence of $V_{xc}$ on time becomes weak whereas in the strongly correlated
limit corresponding to small $x$, it becomes more pronounced. This result may
well be quite general and it is in accordance with expectation since in the
weakly correlated regime, a static mean-field potential is expected to be a
good approximation.

Another interesting feature is the discontinuity of 
$V^{xc}$ at 
$t=0$, which is the difference between the particle ($t=0^+$) and the hole ($t=0^-$)
values, reminiscent of the discontinuity in 
the exchange-correlation potential in density functional theory \cite{perdew1982}.

One also notices that the time dependence is dictated by the excitation
energies of the $(N\pm1)$ systems and in general, these excitations include
collective ones. For example, for solids one expects a time-dependent term of
the form $\exp(i\omega_{p}t)$ where $\omega_{p}$ is the plasmon energy.
$V_{xc}$ acts then like an effective external field, exchanging
an energy
$\omega_{p}$ with the system, as illustrated more explicitly
in the next example on the Holstein Hamiltonian. 


\subsection{Holstein Hamiltonian}

A simplified Holstein Hamiltonian describing a coupling between a core electron and a set
of Bosons, such as plasmons or phonons, is given by \cite{langreth1970}
\begin{equation}
    \hat{H} = \varepsilon \hat{c}^\dagger \hat{c}
    +\sum_q \hat{c} \hat{c}^\dagger g_q(\hat{b}_q + \hat{b}_q^\dagger)
    +\sum_q \omega_q \hat{b}_q^\dagger \hat{b}_q,
\end{equation}
where $\varepsilon$ is the core electron energy, $\omega_q$ is the Boson energy of wave vector $q$,
and $\hat{c}$ and $\hat{b}_q$ are respectively the core electron and the Boson operators. This Hamiltonian can be solved analytically and the algebra is simplified if it is assumed that
the Boson is dispersionless with an average energy $\omega_p$. Under this assumption, the
exact solution for the core-electron removal spectra yields \cite{langreth1970}
\begin{equation}
    A(\omega)=\sum_{n=0}^\infty f_n \delta(\omega-\varepsilon-\Delta\varepsilon+n\omega_p),
\end{equation}
where
\begin{equation}
    f_n=\frac{e^{-a}a^n}{n!},\;\;a=\sum_q \left( \frac{g_q}{\omega_p}\right)^2, 
    \;\; \Delta\varepsilon = a\omega_p.
\end{equation}
This exact solution can also be obtained using the cumulant expansion
\cite{bergersen1973,hedin1965,almbladh1983}. 
The hole Green function corresponding to the above spectra is given by
\begin{equation}
    G(t<0) = i \sum_{n=0}^\infty f_n e^{-i(\varepsilon+\Delta\varepsilon-n\omega_p)t}\theta(-t).
\end{equation}
It can be verified that the time-dependent exchange-correlation potential corresponding to
the Holstein Hamiltonian reads
\begin{equation}
    V_{xc}(t<0)= \Delta\varepsilon \left(1-e^{i\omega_p t}\right).
\end{equation}
This expression provides a very simple interpretation: the first term corrects the non-interacting
core-electron energy whereas the second term describes the Bosonic mode interacting
with the core electron, which can exchange not only one but multiple quanta of $\omega_p$ with
the field.
This is precisely what is accomplished by the cumulant expansion
\cite{bergersen1973,hedin1965,almbladh1983}
within the self-energy formulation but in an \emph{ad hoc} manner.

\section{Conclusion}

The problem of calculating the self-energy is recast into the problem of
constructing the exchange-correlation potential arising from the
exchange-correlation hole, which is a physically motivated entity fulfilling
an exact sum rule and some limiting properties. The proposed formalism is in
the spirit of density functional theory, in which the main task is to find
an accurate approximation for the exchange-correlation functional.

There are many aspects to explore and consider. For example, it would be
interesting to investigate the degree of locality of $V_{xc}$, i.e., how
$V_{xc}$ behaves as a function of $|\mathbf{r}^{\prime}-\mathbf{r}|$ and to
study the time dependence of $V_{xc}$ for a number of model systems such as
the electron gas and one-dimensional systems solvable by means of the Bethe
ansatz in order to accumulate clues and guidelines for constructing an
accurate and reliable $V_{xc}$ or $\rho_{xc}$.

\begin{acknowledgments}
Financial support from the Knut and Alice Wallenberg (KAW) 
Foundation and the Swedish Research Council (Vetenskapsrådet, VR) is gratefully acknowledged. 
The author would like to thank Olle Gunnarsson for his thoughtful comments and suggestions.
\end{acknowledgments}


\end{document}